\documentclass[letter]{aa} 

\usepackage{graphicx}
\usepackage{txfonts}
\usepackage{multirow}
\usepackage{float}
\usepackage{natbib}
\bibpunct{(}{)}{;}{a}{}{,}
\usepackage{placeins}
\usepackage{stfloats}
\usepackage[]{hyperref}
\usepackage{xcolor}

\begin{document}

   \title{\texttt{FlopPITy}: enabling self-consistent exoplanet atmospheric retrievals with machine learning.}
    
    \titlerunning{\texttt{FlopPITy}}
   \author{F. Ard\'evol Mart\'inez
          \inst{1,2,3,4}
          \and
          M. Min\inst{2}
          \and
          D. Huppenkothen\inst{2}
          \and
          I. Kamp\inst{1}
          \and
          P. I. Palmer\inst{3,4}
          }

   \institute{Kapteyn Astronomical Institute, University of Groningen, Groningen, The Netherlands\\
              \email{ardevol@astro.rug.nl}
         \and
             Netherlands Space Research Institute (SRON), Leiden, The Netherlands
        \and
            Centre for Exoplanet Science, University of Edinburgh, Edinburgh, UK
        \and
            School of GeoSciences, University of Edinburgh, Edinburgh, UK
             }

   \date{Received XXX; accepted YYY}

 
  \abstract
   {
   Interpreting the observations of exoplanet atmospheres to constrain physical and chemical properties is typically done using Bayesian retrieval techniques. Because these methods require many model computations, a compromise is made between model complexity and run time. Reaching this compromise leads to the simplification of many physical and chemical processes (e.g. parameterised temperature structure).
   } 
   {Here we implement and test sequential neural posterior estimation (SNPE), a machine learning inference algorithm, for exoplanet atmospheric retrievals. The goal is to speed up retrievals so they can be run with more computationally expensive atmospheric models, such as those computing the temperature structure using radiative transfer.}
   {We generate 100 synthetic observations using ARCiS (ARtful Modeling Code for exoplanet Science, an atmospheric modelling code with the flexibility to compute models in varying degrees of complexity) and  perform retrievals on them to test the faithfulness of the SNPE posteriors. The faithfulness quantifies whether the posteriors contain the ground truth as often as we expect. We also generate a synthetic observation of a cool brown dwarf using the self-consistent capabilities of ARCiS and run a retrieval with self-consistent models to showcase the possibilities that SNPE opens.}
   {We find that SNPE provides faithful posteriors and is therefore a reliable tool for exoplanet atmospheric retrievals. We are able to run a self-consistent retrieval of a synthetic brown dwarf spectrum using only 50,000 forward model evaluations. We find that SNPE can speed up retrievals between $\sim2\times$ and $\geq10\times$ depending on the computational load of the forward model, the dimensionality of the observation, and the signal-to-noise ratio of the observation. We make the code publicly available for the community on Github.}
   {}

   \keywords{planets and satellites: atmospheres}

   \maketitle
%


\section{Introduction}

Interpreting exoplanet and brown dwarf observations to  estimate the physical and chemical properties of their atmospheres is typically done using Bayesian inference to find the joint posterior probability distribution of model parameters. This posterior is traditionally found using sequential sampling-based inference methods, most commonly Markov Chain Montecarlo (MCMC) or different nested sampling algorithms, particularly Multinest \citep{Feroz2009MultiNest:Physics}. This is a computationally expensive process, often requiring hundreds of thousands to millions of forward model evaluations to converge. 

The high computational expense essentially limits the use of complex atmospheric models in retrievals, making it necessary to reach a compromise between model complexity and compute time. The model complexity arises from including more realistic physics, e.g. computing self-consistently the temperature structure and cloud formation or including disequilibrium chemistry.
Models that take more than a second to run already push a single retrieval to between a day and two weeks depending on the number of model evaluations needed.
In addition, higher spectral resolution or larger wavelength coverage with new instruments increase the size of the data set, and are often coupled with higher sensitivities e.g.~by JWST\citep{Gardner2006TheTelescope}. Together, these cause inference methods to require more model evaluations to converge, easily reaching $10^{7}$ evaluations based on retrievals in \cite{barrado+23}.

In order to speed up retrievals and enable analyses of more detailed observations with more complex models, machine learning retrieval methods have started to be developed \citep[e.g., ][]{Waldmann2016DREAMINGATMOSPHERES, Zingales2018ExoGAN:Networks, Marquez-Neila2018SupervisedAtmospheres, Cobb_2019, nixon2020ml, Yip2020PeekingRetrievals, ardevol+22,yip+vi+22, vasist23}. 

These previous studies have shown that machine learning can provide constraints compatible with those obtained with Multinest. 
In particular, \cite{ardevol+22} showed the parameter constraints obtained with machine learning retrievals to be extremely reliable. 
However in this previous approach the posterior was approximated by a multivariate Gaussian. Although with good enough data it would be a reasonable assumption, this is not yet the case for exoplanet atmospheres, so the reliability came at the expense of inaccurate posterior shapes. 
To fix this, \cite{vasist23} developed a normalizing flow based retrieval framework that removed the assumption of gaussianity and is able to reproduce the shape of nested sampling posteriors. 

The vast majority of previous approaches have been amortised estimators.  This means that after training, they are suitable to perform inference on observations occupying any region of parameter space. To accomplish this a large enough training set covering the full parameter space in sufficient resolution is required.  As we already made explicit in \cite{ardevol+22} Section 3.4, the need for a large training set leads to inflexibility with regards to the atmospheric model used.

To counter this shortcoming, \cite{yip+vi+22} developed a machine learning retrieval framework based on variational inference and normalising flows that was able to reproduce accurately nested sampling posteriors using <$10\%$ of the forward models while retaining the same flexibility.
However, this approach requires the atmospheric model to be differentiable and so one would need to either use \verb|Diff|-$\tau$ \citep[the differentiable model developed by][]{yip+vi+22}, or develop new differentiable models, limiting its usability.


Here we present \verb|FlopPITy| (normalising FLOw exoPlanet Parameter Inference ToolkYt), a machine learning retrieval framework based on neural spline flows \citep{neural_spline_flows} and sequential neural posterior estimation \citep[SNPE,][]{greenberg+19}. 
This approach retains the flexibility of sampling-based methods while requiring only a fraction of the forward model evaluations. 
Additionally, it works with any atmospheric modelling code
without the need to rewrite it or adapt it. 

In this letter, we first describe briefly the machine learning approach we use in Section \ref{sec:snpec}. In Section \ref{sec:mock_retrievals} we apply \verb|FlopPITy| to synthetic observations to test and showcase its performance.  Section \ref{sec:discussion} discusses the implications of these results and the advantages and shortcomings of our method. Finally, in Section \ref{sec:conclusion} we present our conclusions.

\section{(Sequential) Neural Posterior Estimation}\label{sec:snpec}

\texttt{FlopPITy}\footnote{ \texttt{FlopPITy}  can be accessed \href{https://github.com/franciscoardevolmartinez/FlopPITy}{here.}} (normalising FLOw exoPlanet Parameter Inference ToolkYt) is a new tool for exoplanet atmospheric retrievals that uses sequential neural posterior estimation. In particular, we use Automatic Posterior Transformation \citep[APT or SNPE-C, ] []{greenberg+19} as implemented in the python package SBI \citep{sbi}. 
SNPE-C belongs to a larger group of likelihood-free inference methods, which are useful when the likelihood function is intractable. In our case the likelihood can be calculated, so the advantage of SNPE is the speed-up that the use of machine learning provides.
SNPE-C approximates the posterior distribution $p(\theta|x)$ with the distribution $q_{F(x,\phi)}(\theta|x)$, where $q$ is a density family\footnote{A density family is a group of probability distributions described by the same parameters. Examples of density families are: the Gaussian distribution, a Gaussian mixture with N components, or the binomial distribution.} and $F$ is a neural network with weights $\phi$. In this work we use normalizing flows \citep{normalizing_flows} for our posterior estimator $q_F$. See Appendix A in \cite{vasist23} for a brief overview on normalizing flows. More specifically, we use neural spline flows \citep{neural_spline_flows}, for which a more palatable explanation can be found in \cite{Green_2021}. 

\subsection{Multi-round training}\label{sec:multi-round}
Amortised estimators such as those presented in \cite{ardevol+22} and \cite{vasist23} are incredibly convenient as once trained, inference can be carried out almost instantly for any observation. However, they are relatively inflexible. Observational details and data processing choices (e.g.~wavelength range used, spectral resolution, noise properties) are fixed during training, which means that for real world usage the estimator needs to be retrained with the observational properties of each observation. 
The main source of inflexibility comes from the need to pre-compute large training sets covering the whole prior parameter space, limiting the use cases for the trained estimator. This is because simple changes like adding extra chemical species to the models imply computing a whole new training set. Additionally this large upfront computational cost, although lower than would be required for Multinest, can still be unfeasible for more complex models.

Here we present a non-amortised approach that is as flexible as traditional sampling-based retrievals. Like the latter, it requires computing new forward models for every retrieval. However it needs only a fraction of them compared to nested sampling, resulting in a significant speed-up.

\begin{figure}
    \centering
    \includegraphics[width=0.5\textwidth]{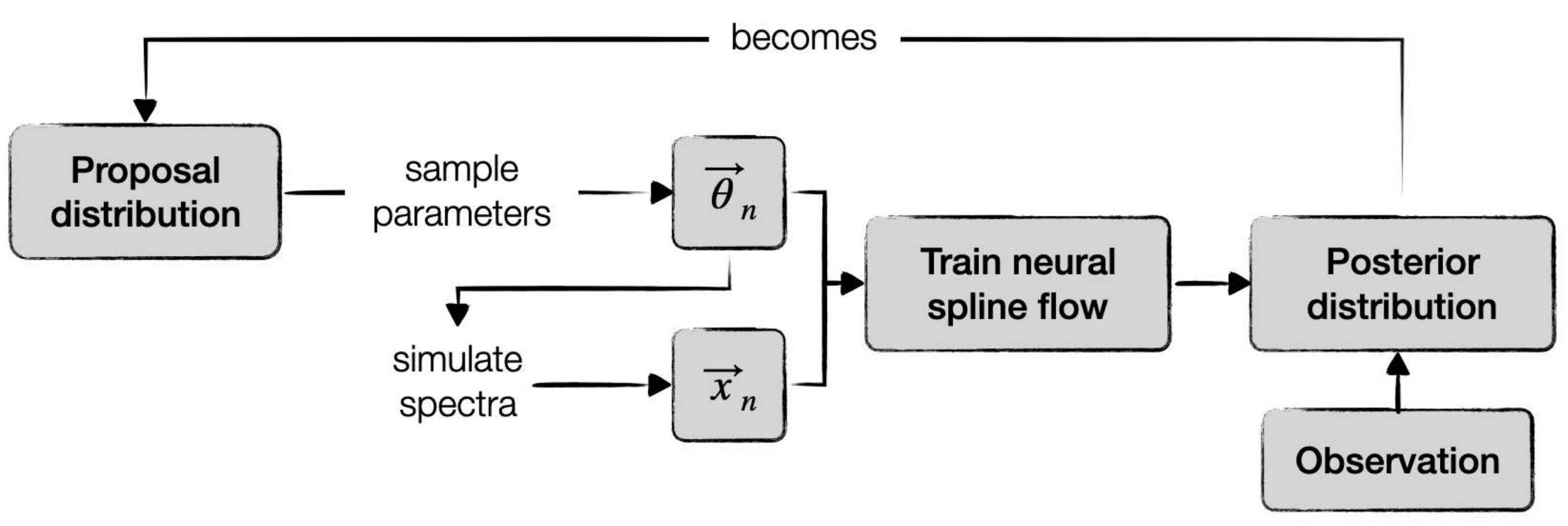}
    \caption{Schematic representation of the iterative training of a neural spline flow. In the very first iteration, the proposal distribution is the prior.}
    \label{fig:diagram}
\end{figure}

We use SNPE-C with multi-round training as described in \cite{greenberg+19}. The way it works is as follows: initially, $N$ parameter vectors $\vec{\theta}_i$ are drawn from the prior, and corresponding forward models $\Vec{x}_i$ are computed. These are used to train the estimator and obtain a first approximation to the posterior. From this posterior, we draw another $N$ samples and compute the corresponding forward models, which we use to improve the training and obtain an improved estimate for the posterior distribution. Figure \ref{fig:diagram} illustrates how the method works. After the first round of training the estimator is no longer amortised but is only suitable for the specific observation we are analysing. This process is repeated for a set number of rounds. Unfortunately, because after the first round we are not sampling from the prior $p(\vec{\theta})$ but rather a so-called proposal distribution $\Tilde{p}(\vec{\theta})$, the resulting distribution is no longer the true posterior $p(\vec{\theta}|\vec{x})$ but rather a so-called proposal posterior:

\begin{equation}\label{eq:tildeP}
    \Tilde{p}(\vec{\theta}|\vec{x}) = p(\vec{\theta}|\vec{x})\frac{\Tilde{p}(\vec{\theta})p(\vec{x})}{p(\vec{\theta})\Tilde{p}(\vec{x})}
\end{equation}
where 

\begin{equation}
    \Tilde{p}(\vec{x})=\int_{\vec{\theta}}\Tilde{p}(\vec{\theta})p(\vec{x}|\vec{\theta}).
\end{equation}

SNPE-C automatically transforms between estimates of the true posterior and the proposal posterior, making it easy to sample the estimated true posterior.
The interested reader can consult \cite{greenberg+19} for the details of how this is accomplished. 

\color{black}{Since SNPE-C uses $q_{F(x,\phi)}(\theta)$ to approximate the posterior, from eq. \ref{eq:tildeP} we can approximate the proposal posterior as $\Tilde{q}_{F(x,\phi)}(\theta)\propto q_{F(x,\phi)}(\theta)\Tilde{p}(\theta)/p(\theta)$.
We train the network by minimizing the loss function:
\begin{equation}
    \mathcal{L(\phi)}=-\sum_{j=0}^N \log \Tilde{q}_{x,\phi}(\theta_j)
\end{equation}
This yields $q_{F(x,\phi)}(\theta)\rightarrow p(x|\theta)$ and $\tilde{q}_{F(x,\phi)}(\theta)\rightarrow \tilde{p}(x|\theta)$ as $N\rightarrow \infty$ \citep{papamakarios+16}.
The proposal prior is then defined as $\tilde p(\theta) = q_{F(x_0,\phi)}(\theta)$, where $x_0$ is the data.
}

\color{black}

This sequential approach is not without caveats, as it can provide overconfident posteriors \citep{overconfident_snpec}. However we have not found that to be the case and this is discussed in more detail in Section \ref{sec:mock_retrievals}. 

\color{black}{The observational error is accounted for by adding noise to the simulated spectra. Training is done on noisy samples $x_i'=x_i+\varepsilon$ with $\varepsilon\in\mathcal{N}(0,\sigma_0)$, where $x_i$ are the simulated spectra, $\mathcal{N}(\mu,\sigma)$ is the normal distribution, and $\sigma_0$ the observed noise. Because the network has already learnt the noise, at the time of inference only $x_0$ is passed.}

 \color{black}
\section{Mock retrievals}\label{sec:mock_retrievals}

We test \texttt{FlopPITy} on two scenarios. First, we run retrievals using a simple and fast atmospheric model to test its reliability. Second, we run a retrieval using a complex, computationally expensive atmospheric model to showcase the science cases that this methods enables.

\subsection{Bulk retrievals}\label{sec:bulk_retrievals}

We perform retrievals on a hundred synthetic NIRSpec PRISM spectra to test the faithfulness of our method. For the purpose of illustration, we use a relatively simple (computationally fast) atmospheric model to minimise the computational load. 

The synthetic observations are generated using ARCiS \citep{arcis} with an isothermal temperature structure $T(p) = T_0$, free H$_2$O and CO$_2$ abundances, and the radius $R_P$ and $\log{g}$ of the planet. The parameter ranges from which the synthetic spectra were sampled are the same as the prior ranges used for the retrievals, and are shown in Table \ref{tab:priors_bulk}. The spectroscopic channels and observational noise are taken from the FIREFLy reduction in \cite{rustamkulov+23}.

To perform the retrievals, we train \texttt{FlopPITy} in 20 rounds using 1,000 simulations in each round. Such a high number of rounds is not necessary but this allows us to check if training for more rounds than necessary leads to overconfidence. We also perform Multinest retrievals on the simulated observations to have a baseline to compare against.


\begin{figure}[t]
    \centering
    \includegraphics[width=0.49\textwidth]{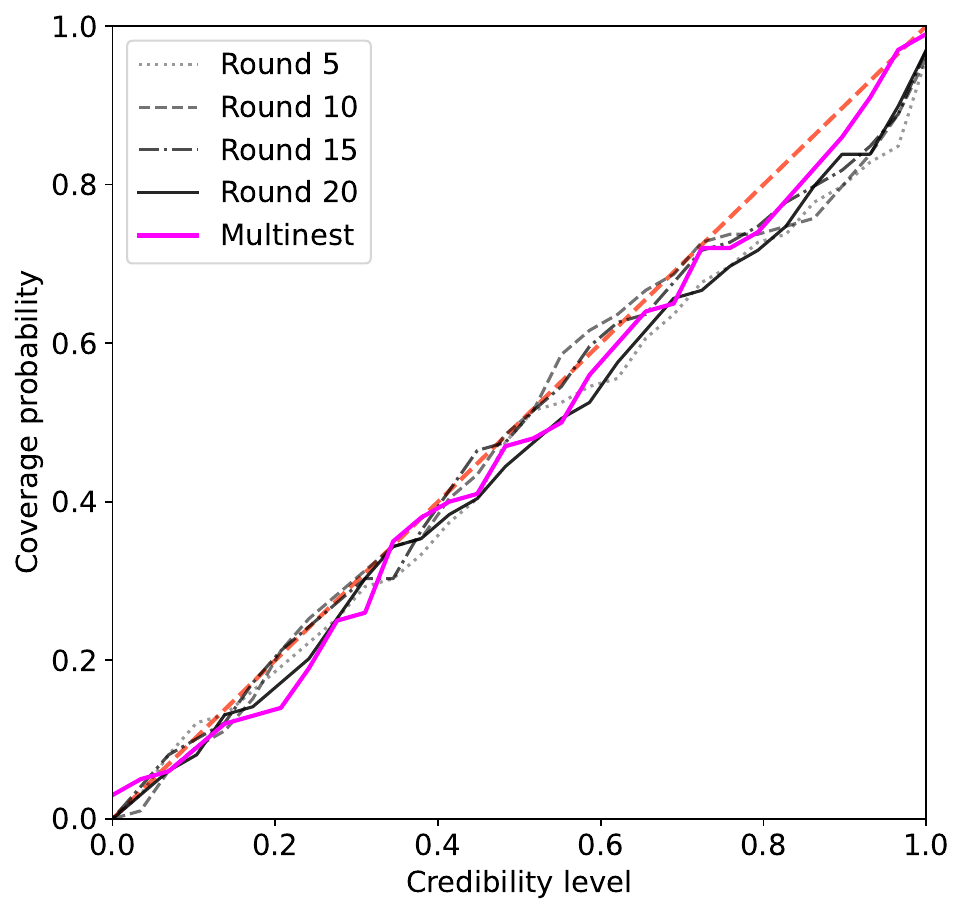}
    \caption{Coverage probability of SNPE posteriors at different training rounds compared to Multinest. The red dashed line denotes the 1:1 line. 
    All the lines are close to the diagonal, indicating that both the Multinest and SNPE posteriors are faithful. Importantly, the latter remain faithful at every round.
    }
    \label{fig:coverage_plot}
\end{figure}

We want to check that the posteriors produced by our method are correctly estimated and are not too broad or too narrow. We do this by calculating the expected coverage probability following \cite{overconfident_snpec} and \cite{vasist23}.  
The coverage probability is the probability of a certain confidence region containing the ground truth. If the posteriors were correctly estimated, a region of the posterior with a fraction ($1-\alpha$)\% of the probability would contain the ground truth ($1-\alpha$)\% of the time. Figure \ref{fig:coverage_plot} shows the coverage probability for the posteriors produced at each round. Its interpretation is simple: if the coverage probability is below the diagonal, the posteriors are overconfident (too narrow), whereas if the coverage probability is above, posteriors are underconfident (too wide). 

Although the curves in Fig. \ref{fig:coverage_plot} are a bit noisy due to computational feasibility limiting the number of mock retrievals, they follow the diagonal closely at each training round, showing a performance on par with that of Multinest. Crucially, Fig. \ref{fig:coverage_plot} shows that the posteriors do not become increasingly overconfident with subsequent training iterations, but are reliable at each step. 
\color{black}{Figure \ref{fig:coverage_plot} seems to indicate that at high credibility levels, \texttt{FlopPITy} is slightly  overconfident when compared to Multinest. A larger number of mock retrievals would need to be run to be able to ascertain whether it is a real effect or just an artifact. If real, it would indicate that the probability in the wings of the posteriors is underestimated and therefore care should be taken not to overinterpret them.}

\color{black}

\begin{figure*}[t]
    \centering
    \includegraphics[width=0.95\textwidth]{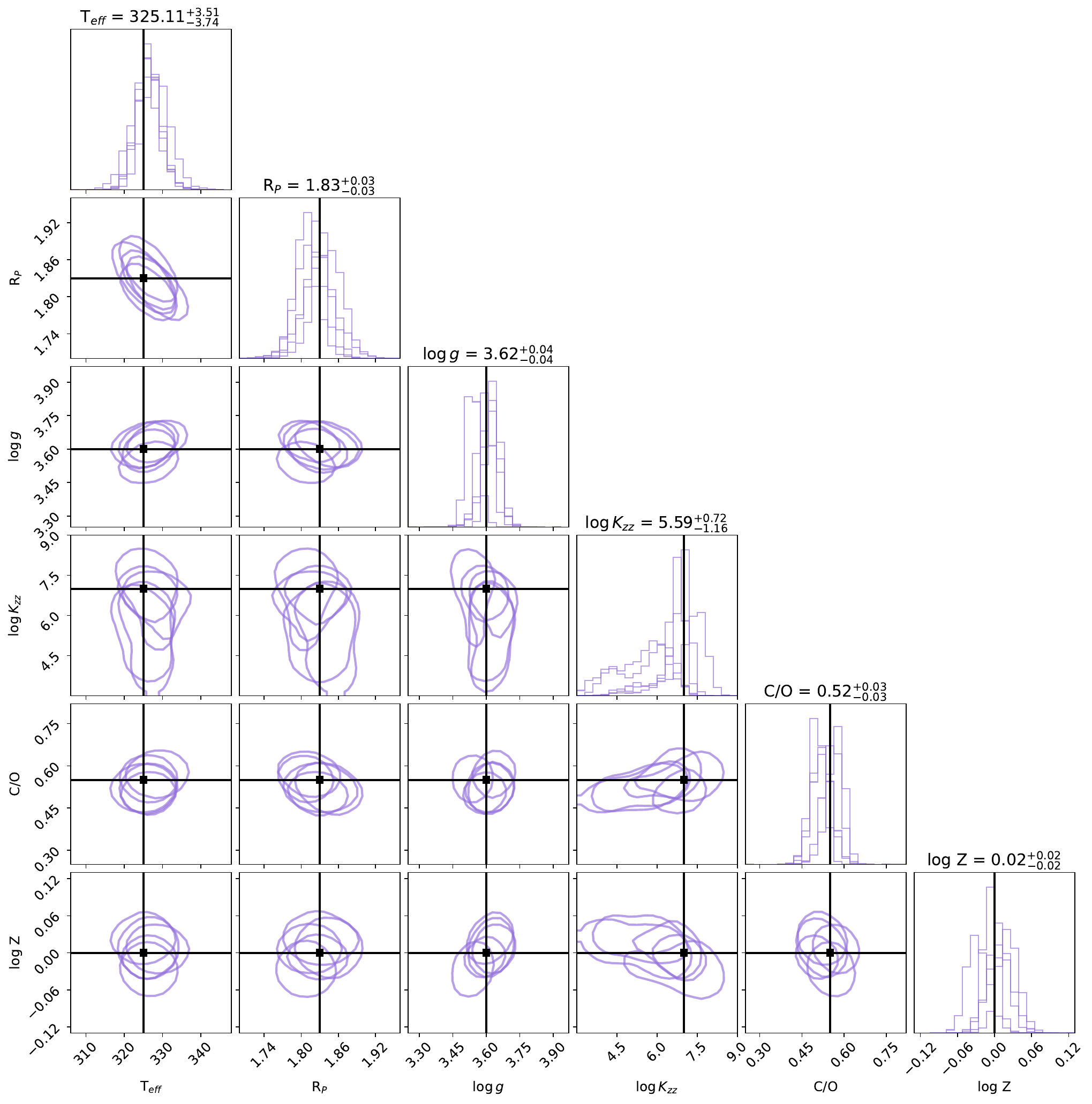}
    \caption{Posteriors for five noise realizations of a synthetic WISE 1828+2650 spectrum. The input parameters of the synthetic spectrum are denoted by the black lines and are: T$_{\text{eff}}=325$\,K, R$_\text{P}=1.83$\,R$_\text{J}$, $\log{g}=3.6$, $\log{K_{\text{zz}}}=7$, C/O$=0.55$ and $\log{Z}=0$. To keep the figure readable, we show only the 2$\sigma$ contours. The quantiles shown in the titles correspond to one of the noisy realisations.} 
    \label{fig:selfcon_corner}
\end{figure*}

\subsection{Self-consistent retrieval}\label{sec:selfcon_retrieval}

To showcase the real power of \verb|FlopPITy|, we perform a retrieval with a self-consistent setup on a simulated observation of a brown dwarf. 
We generate the synthetic observation using the self-consistent capabilities of ARCiS. We assume a 1D (cloud free) atmosphere in radiative-convective equilibrium (with an effective temperature $T_{\text{eff}}$) as well as thermochemical equilibrium computed from the carbon to oxygen ratio C/O and metallicity $Z$, and include chemical disequilibrium due to vertical mixing \citep{yui+min}. The vertical mixing is parameterised by the eddy diffusion coefficient $K_{\text{zz}}$. 
We use the best fitting parameters for the cool Y dwarf WISE 1828+2650 \citep{j1828_nircam}.  
We generate a spectrum with $R=1,000$ in the wavelength range $5 - 18\,\mu$m. We assume an error on the flux of $10^{-5}$\,Jy (corresponding to an SNR between $\sim$~0 and $\sim$~40 depending on the wavelength), which is representative of what can be achieved in most of the MIRI MRS wavelength range when binned down to $R=1,000$ \citep{barrado+23}.
The model parameters and their priors are shown in Table \ref{tab:priors_sc}.
We do not show a comparison to a Multinest retrieval as it is not computationally feasible.

We train \verb|FlopPITy| in ten rounds, using 5,000 forward models in each round.
We use a larger simulation budget than in the previous case  because the mapping from atmospheric parameters to spectra is more complex. 
In the simple model case, changes in parameters directly affect the spectrum,  e.g. a higher molecular abundance increases the amplitude of the features in the spectrum.
For the self-consistent model this is no longer the case.
For example, the molecular abundances (which influence the amplitude of the spectral features) depend on the internal temperature, the carbon-to-oxygen ratio, the metallicity, and the vertical diffusion. 

To account for possible biases due to the noise realization of an individual observation, we generate five different noisy observations from the simulated spectrum and aggregate the posteriors retrieved.
The corresponding corner plot can bee seen in Fig. \ref{fig:selfcon_corner}, with all parameters being faithfully recovered.

Figure \ref{fig:selfcon_corner} shows that for the SNR considered, we can get errors of $\sim1$\% on crucial evolutionary properties, such as $T_{\text{eff}}$, $R$ and $\log g$. This is only the case when the model perfectly reproduces the observation. For real data, this will virtually never be the case as there will always be physical/chemical processes not taken into account by the model. In this case it is unknown what level of precision and bias one should expect, and even to which extent the retrievals remain reliable. There might also be differences in performance between \texttt{FlopPITy} and Multinest, making one of them a better option. These questions will be explored in future work.

\section{Discussion}\label{sec:discussion}

We have shown that \verb|FlopPITy|, an implementation of SNPE-C with neural spline flows, is a reliable method for exoplanet and brown dwarf atmospheric retrievals and it allows us to perform retrievals using forward models that are too slow for traditional sampling-based retrievals. 
In particular, the retrieval performed on the simulated brown dwarf spectrum  (Section \ref{sec:selfcon_retrieval}) used 50,000 simulations in total, taking $\sim$18\,h to complete on a 2020 M1 MacBook Pro. 
This time was split in $\sim$15\,h for the computation of the simulations and $\sim$3\,h for training. 
Based on the retrievals ran in \cite{barrado+23}, we can estimate that the same retrieval with Multinest would require between 500,000 and 1,500,000 model evaluations to converge.
Since each forward model requires $\sim$3\,s to run, such a retrieval would take at least $\sim$ 20 days to converge. 

As a comparison, the simple retrievals took $\sim 3$\,h each in a cluster with 48 AMD Opteron 6348 processors, out of which $\sim1$\,h 25\,min were used for training. Their Multinest counterparts typically needed between 20,000 and 60,000 forward models to converge, taking between 3 and 8 hours. Higher SNR observations typically require more forward models for Multinest to converge. The lower (to no) computational advantage of \verb|FlopPITy| over Multinest in this case is due to two factors. First, having a simpler dataset (with a factor $\sim6$ lower spectral resolution) results in Multinest requiring fewer model evaluations to converge. Second, using a faster forward model ($\sim$0.5\,s versus $\sim$3\,s) causes training to represent a significant fraction of the time needed for the \verb|FlopPITy| retrieval.

This means that for simpler datasets and faster forward models, our method does not provide a substantial speed up compared to Multinest, and for very simple datasets (e.g. HST WFC3 spectra), Multinest might even be preferred. Conversely, \texttt{FlopPITy} enables analyses of complex datasets with computationally costly atmospheric models that would otherwise not be feasible with sampling-based retrieval methods.

Additionally, Multinest has the disadvantage that models need to be computed sequentially, so it can not be parallelised
\footnote{A small level of parallelization can be achieved by simultaneously drawing $N$ points at each nested sampling iteration, with $1/N$ being an estimate of the sampling efficiency.}. SNPE does not have this limitation, and the computation of the forward models can be spread over multiple CPUs, further speeding up retrievals.

The sequential approach presented here is not universally preferable to amortised approaches such as the one in \cite{vasist23}. 
In particular, if the goal is to perform retrievals on a large number of spectra using the same atmospheric model \citep[as will undoubtedly be the case for future missions such as ARIEL, ][]{ariel}, an amortised approach will be computationally more efficient. 
However for the exploration of a single dataset, our sequential approach is more appropriate as the extra flexibility  allows to more easily try and compare different atmospheric models with different assumptions.

The main drawback of \verb|FlopPITy| is the need to choose how many rounds to train for, and how many training examples to use per round. 
{ Fortunately, as we have shown in Section \ref{sec:bulk_retrievals}, the posteriors do not become overconfident by training for too many rounds. Nevertheless training for more rounds than necessary represents an unnecessary computational expense that we would like to avoid. } 
A way to gauge the convergence of the retrieval is to compare the posteriors at consecutive rounds, and train for a few more rounds if there is still significant variation. 
The $1\sigma$, $2\sigma$ and $3\sigma$ envelopes of the retrieved spectra can also be used to ensure that the posterior is not over-dispersed. 
Regarding the amount of training examples, ideally one would choose as many as computationally affordable. 
Due to the stochastic nature of the random sampling of parameter vectors in the prior and proposal distributions, using too few examples will result in an uneven coverage of the parameter space, which could have a negative effect on the results. 

Finally, it is still unclear how \verb|FlopPITy| or nested sampling respond to adversarial examples. These are observations with features that are not well reproduced by the model, which we coined `uncomfortable retrievals' in \cite{ardevol+22}. There we showed that  machine learning  was more reliable than Multinest when the observations were not well reproduced by the underlying atmospheric model. However, this came at the expense of very broad posteriors, so the information gain over the prior was limited (which is still preferred to biased posteriors). In future work we will explore if \verb|FlopPITy| can remain reliable in this scenario while being more informative than the machine learning retrievals in \cite{ardevol+22}, or if it instead behaves more similarly to Multinest.

\section{Conclusions}\label{sec:conclusion}

In this letter we present \verb|FlopPITy|, a new machine learning retrieval tool that uses normalizing flows and multi-round training, and we show that it works reliably.

In contrast with previous machine learning retrieval methods, this method retains the flexibility of sampling-based methods and is applicable to any atmospheric model, requiring only a fraction of the forward models needed by the latter, with the exact fraction depending on the specific observation and model used. 
When performing individual retrievals, the sequential approach presented here requires fewer models to train on than amortised approaches, as the computational effort is focused in the relevant regions of parameter space. However, once trained, amortised estimators are able to perform retrievals almost instantly and are therefore better suited for the analysis of large datasets with the same forward model.

Our method enables retrievals of high quality observations, such as those provided by JWST, with computationally costly forward models, e.g. self-consistent temperature structure or cloud formation. 
This reduces the need to simplify atmospheric models to speed them up for retrievals, although of course it does not eliminate it completely, as still tens of thousands of forward models need to be computed in a reasonable time frame. 
Additionally, unlike with sampling-based retrievals, the computation of these models can be parallelised over multiple CPUs, further speeding up a retrieval.

This work highlights the avenues that machine learning is opening for characterizing exoplanets and expands the suite of existing machine learning retrieval methods and their use cases.

\begin{acknowledgements}

      We would like to thank the Center for Information Technology of the University of Groningen for their support and for providing access to the Peregrine high performance computing cluster. This research has made use of NASA’s Astrophysics Data System. This project has received funding from the European Union’s Horizon 2020 research and innovation programme under the Marie Sklodowska-Curie grant agreement No. 860470.\\

      We would like to thank Malavika Vasist for fruitful discussions. We are thankful to Beatriz Campos Estrada, whose comments helped improve the manuscript. We are also grateful to Daniel Talav\'an Vega, who came up with the name \texttt{FlopPITy} over half a decade ago.\\

      PIP acknowledges funding from the STFC consolidator grant \#ST/V000594/1.
      
\end{acknowledgements}

%
\bibliographystyle{aa} 
\bibliography{main.bib}
%

\begin{appendix}\label{app:priors}

\section{Retrieval priors}

The tables below contain the priors used in the retrievals shown in the text.

\begin{table}[h]
    \centering
    \begin{tabular}{c|c|c|c}
        \hline
         Parameter & Prior & Units & Shape\\
         \hline
         \hline
         $T$ & 10 - 3000&K&Log-uniform\\
         VMR(H$_2$O) & 10$^{-12}$ - 10$^{0}$&&Log-uniform\\
         VMR(CO$_2$) & 10$^{-12}$ - 10$^{0}$&&Log-uniform\\
         $R_P$ & 0.1 - 1.9&R$_J$&Uniform\\
         $\log{g}$ & 2 - 4& cm s$^{-2}$ for $g$&Uniform\\
         \hline
         
    \end{tabular} 
    \vspace{0.3cm}
    \caption{Bulk retrieval priors.}
    \label{tab:priors_bulk}
\end{table}


\begin{table}[h]
    \centering
    \begin{tabular}{c|c|c|c}
        \hline
         Parameter & Prior & Units & Shape\\
         \hline
         \hline
         $T_{eff}$ & 10 - 3000&K&Log-uniform\\
         $R$ & 1 - 4&R$_J$&Uniform\\
         $\log{g}$ & 2.8 - 6& cm s$^{-2}$ for $g$&Uniform\\
         $K_{zz}$ & 10$^3$ - 10$^{12}$& cm$^2$ s$^{-1}$& Log-uniform\\
         C/O & 0.01 - 2&&Uniform\\
         ${Z}$ & 10$^{-3}$ - 10$^3$&$Z_{\odot}$&Log-uniform\\
         \hline
    \end{tabular}     
    \vspace{0.3cm}

    \caption{Self-consistent retrieval priors.}
    \label{tab:priors_sc}
\end{table}

\section{When inference fails}\label{app:mock_retrievals}

\begin{figure}
    \centering
    \includegraphics[width=0.5\textwidth]{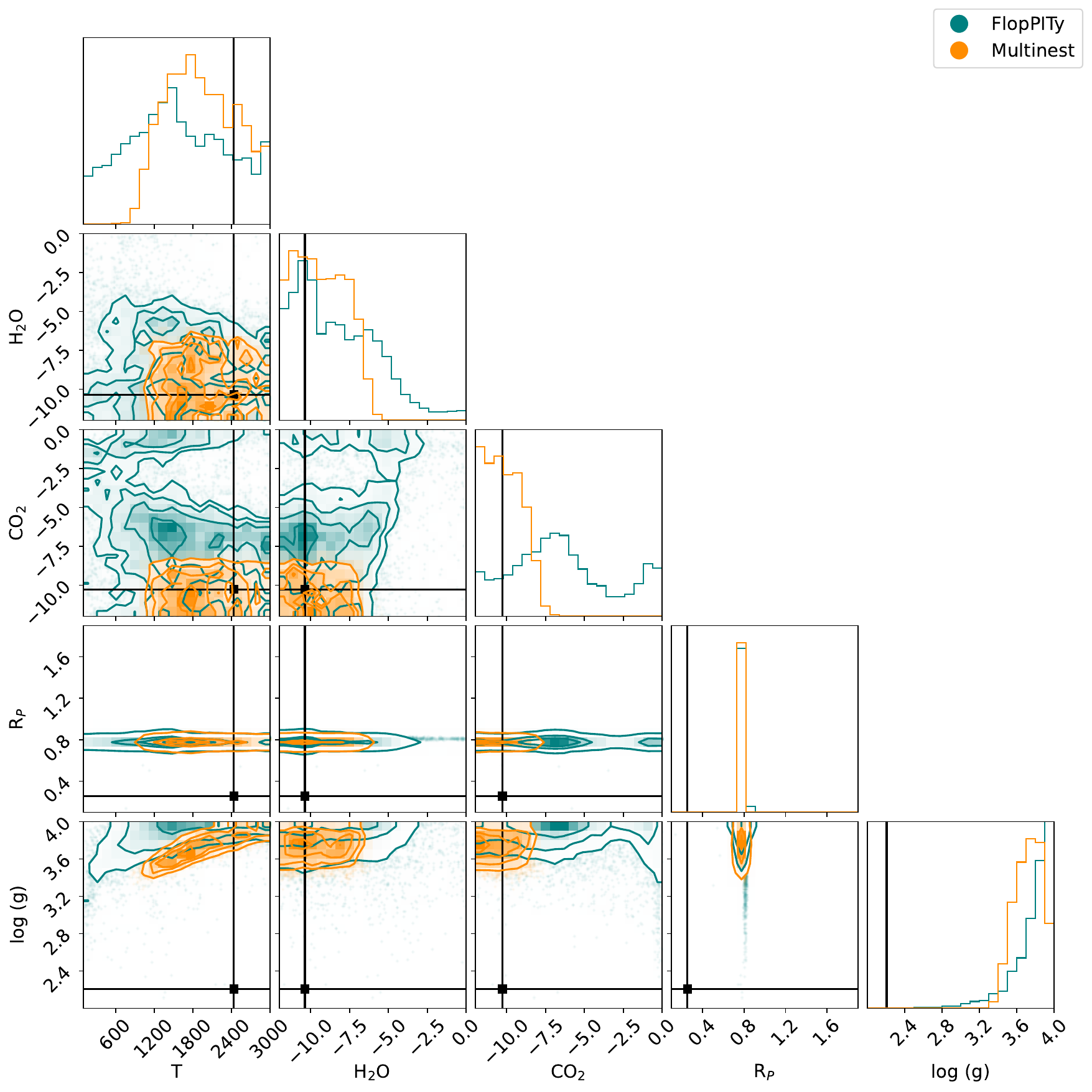}
    \includegraphics[width=0.5\textwidth]{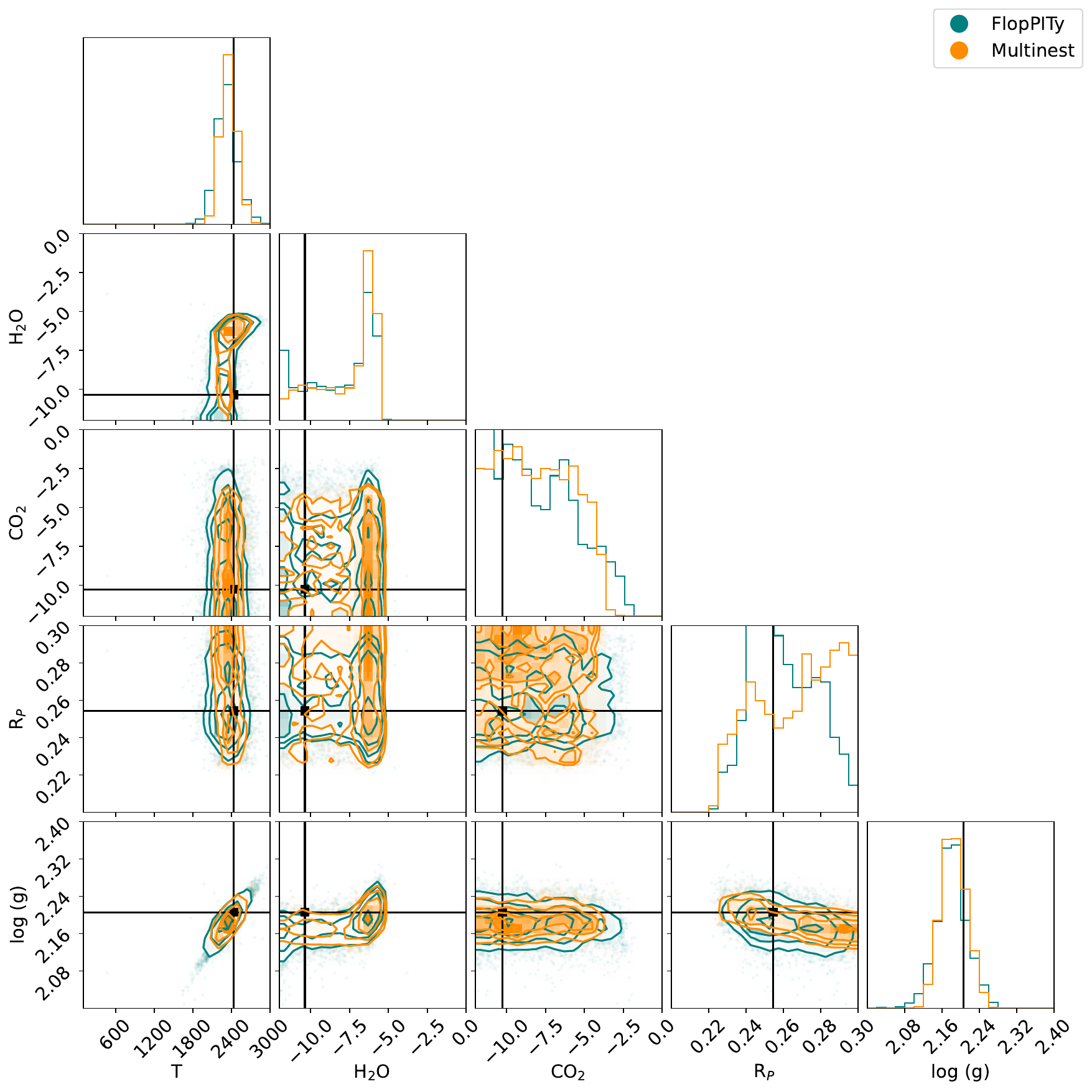}
    \caption{Example corner plots for a case with a very narrow posterior around the ground truth for retrievals run with broad (\textit{top}) and tight (\textit{bottom}) priors.}
    \label{fig:3_broad}
\end{figure}

Both Multinest and  \verb|FlopPITy| can fail in certain situations. Here we show one example from the bulk retrievals in \ref{sec:bulk_retrievals} where the posterior has a narrow mode around the ground truth which is completely missed by both retrieval methods, as visible in Fig. \ref{fig:3_broad}.

In a scenario more representative of real world retrievals, the priors on $R_P$ and $\log g$ would be significantly narrower, as these quantities can be measured from the white light curve and radial velocity, respectively. When we redo the retrievals using such narrow priors, we see that both methods find the right mode (Fig. \ref{fig:3_broad} \textit{bottom}).

\begin{figure}
    \centering
    \includegraphics[width=0.5\textwidth]{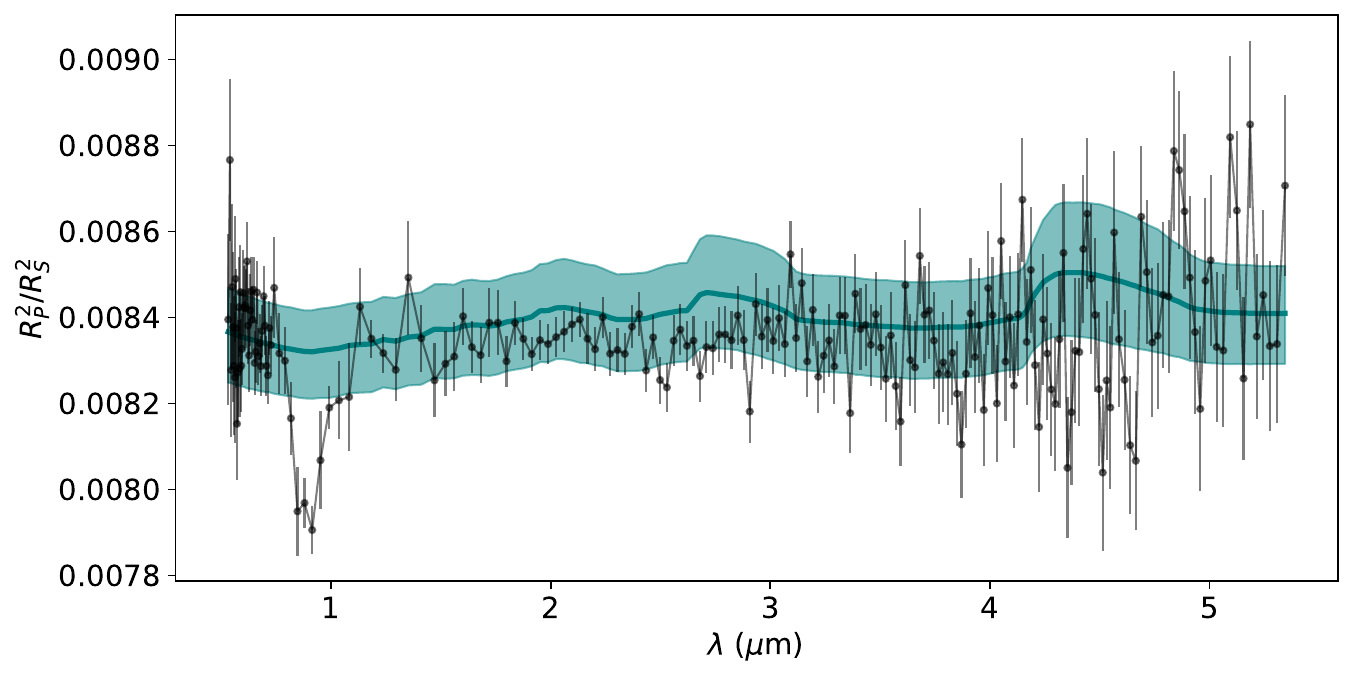}
    \includegraphics[width=0.5\textwidth]{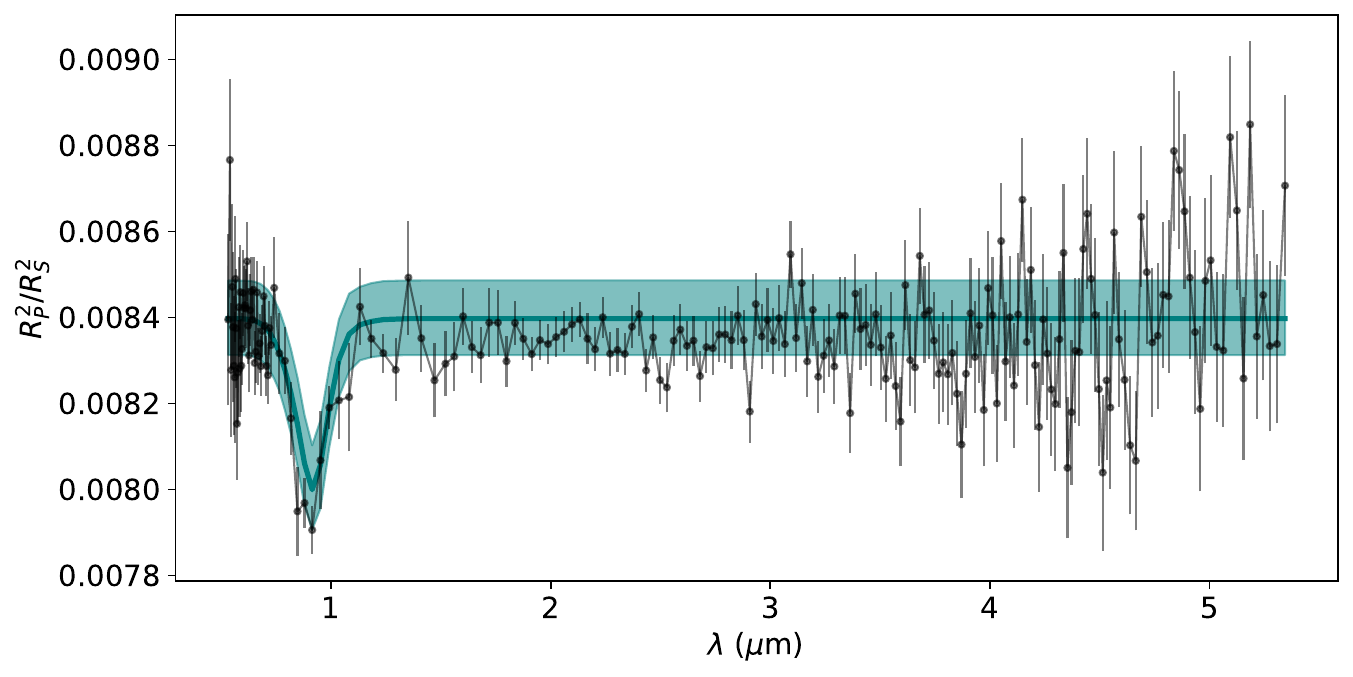}
    \caption{1$\sigma$ contours of the retrieved spectra for the case with broad (\textit{top}) and tight (\textit{bottom}) priors.}
    \label{fig:retrieved_spectra}
\end{figure}

An essential diagnosis for any inference method is to run simulations for samples drawn from the posterior and compare them to the data. As can be seen in Fig. \ref{fig:retrieved_spectra}, this would make it evident that the retrieval with broad priors is not finding the right solution.

This example does not correspond to a physically plausible atmosphere. Since the spectra are sampled randomly from a large parameter space, not all combinations correspond to realistic atmospheres (in this particular case, the high temperature and low gravity causes most of the atmosphere to escape). This is not an issue since we are only interested in seeing if the retrieval methods can find the parameters that gave rise to the simulation. The case presented here is simply a nice illustration of a possible failure mode for Multinest and \verb|FlopPITy|.

\color{black}
\section{Training hyperparameters}\label{hypers}

{Table \ref{tab:hyperparameters} shows the training hyperparameters as well as the neural spline flow structure used.}
\color{black}{
\begin{table}[h]
    \centering
    {\small\begin{tabular}{c|c|c}
        \hline
         \multirow{2}{*}{{Training hyperparams.}}&{Batch size} & {50}\\
         &{Learning rate} & $5\cdot10^{-4}$\\
         \hline
         \multirow{5}{*}{{Neural spline flow}}& {Bins} & {10}\\
         &{Transforms} & {15}\\
         &{Network architecture} & {Residual network}\\
         &{Blocks} & {2}\\
         &{Hidden units} & {50}\\
         \hline
    \end{tabular} }
    \vspace{0.3cm}
    {\color{black}\caption{{Technical details of our implementation.}}}
    \label{tab:hyperparameters}
\end{table}
}
\end{appendix}

\end{document}